# Towards carbon neutral scientific societies:
# A case study with the International Adsorption Society

Anne Streb,[1,2] Ryan Lively,[3] Philip Llewellyn,[4] Akihiko Matsumoto,[5]
Marco Mazzotti,[1] Ronny Pini,[6] Benoit Coasne[7,*]

[1] *Separation Processes Laboratory, Department of Mechanical and Process Engineering, ETH Zurich, Zurich 8092, Switzerland*

[2] *Climeworks AG, Zurich 8050, Switzerland*

[3] *School of Chemical & Biomolecular Engineering, Georgia Institute of Technology, Atlanta, GA 30332, USA*

[4] *TotalEnergies, OneTech, CSTJF-Pau, France*

[5] *Department of Applied Chemistry and Life Science, Toyohashi University of Technology, 1-1, Hibarigaoka, Tempaku-cho, Toyohashi, Aichi, 441-8580, Japan*

[6] *Department of Chemical Engineering, Imperial College London, London, UK*

[7] *Univ. Grenoble Alpes, CNRS, LIPhy, France*

* To whom correspondence should be addressed: benoit.coasne@univ-grenoble-alpes.fr

**With increasing concerns over climate change, scientists must imperatively acknowledge their share in $CO_2$ emissions.[1,2] Considering the large emissions associated with scientific traveling – especially international conferences – initiatives to mitigate such impact are blooming.[2-4] With the COVID-19 pandemic shattering our notion of private/professional interactions,[5-7] the moment should be seized to reinvent science conferences and collaborations with a model respectful of the environment. Yet, despite efforts to reduce the footprint of conferences, there is a lack of a robust approach based on reliable numbers (emissions, carbon offsetting/removals, etc.) to accompany this shift of paradigm. Here, considering a representative scientific society, the International Adsorption Society,[8] we report on a case study of the problem: making conferences carbon neutral while respecting the needs of scientists. We first provide a quantitative analysis of the $CO_2$ emissions for the IAS conference in 2022 related to accommodation, catering, flights, etc. Second, we conduct two surveys probing our community view on the carbon footprint of our activities. These surveys mirror each other, and were distributed two years before and in the aftermath of our triennial conference (also corresponding to pre/post COVID times). By combining the different parts, we propose ambitious recommendations to shape the future of conferences.**

As the Intergovernmental Panel on Climate Change (IPCC) reports increasingly pessimistic scenarios for Earth's surface temperature by the end of the century[9], there is an increasing drive to drastically decrease our greenhouse gas emissions – both at the individual and collective levels. For many of us, such efforts already translate into small practical climate-positive actions in our daily life. Moreover, the fact that emissions from professional activities can largely surpass those from the private sphere has resulted in action to decrease $CO_2$ emissions in all fields relevant to the primary, secondary, tertiary, and quaternary sectors. For quite some time, many professional sectors have considered themselves as role models owing to the limited environmental impact of their activities. This is the case of the science and research field, which has considered its activities as almost carbon neutral, in stark contrast to the industrial and transportation sectors. In the last few years, by analyzing their carbon footprint, many academic and industrial scientific communities have recognized the severe impact of science-related traveling to attend conferences and to visit collaborators.[10-13] For some of these communities, acknowledging such negative impact is perceived as paradoxical, because their activities are specifically devoted to developing technologies that mitigate climate change and reduce carbon dioxide emissions. This is the case of the International Adsorption Society (IAS), which serves in the present paper as the model for a "real case study" to identify options to make carbon neutral scientific societies compatible with cutting-edge innovative research. With its activities centered on adsorption technologies to design processes for environmental, health and energy applications, the IAS is at the forefront of research and development on carbon capture and storage. Our international society, which gathers a few hundred researchers and engineers both from academia and industry, is strongly committed to reducing its carbon footprint as witnessed by the appointment of an IAS working group on carbon neutrality as early as 2019.

While evidence exists on the excessive carbon footprint of scientific conferences, solving the problem of such large $CO_2$ emissions cannot be as simple as making all scientific conferences on-line or continuing business as usual while financially compensating through available carbon offsets. The solution to this complex equation lies in finding the balance so as the undisputable benefits of scientific interactions are weighed against their associated $CO_2$ emissions. On the one hand, any optimal, viable solution needs to accurately account for the carbon footprint of scientific interactions (e.g. conferences, collaborations, visiting programs). On the other hand, it must consider the benefits of such exchanges on the quality of scientific outcomes. One example is the stringent need for the generation of the younger scientists to meet physically with each other and with their more senior colleagues to form efficient professional networks for the next decades.

The quantification of the carbon footprint of scientific and engineering research needs to rely on a robust and transparent strategy. Unfortunately, there is a lack of consolidated datasets and best practices, despite several interesting initiatives having already originated in other communities. To fill this gap, this contribution proposes a rational and transparent approach based on the following two-step strategy. First, considering that the IAS carbon footprint is largely dominated by its international conference held every three years, we present a robust $CO_2$ emission assessment based on accurate attendance numbers and cross-checking of data from different reliable sources. Second, we analyze the results from two surveys that were distributed to assess the opinion of the IAS community regarding its carbon footprint and conferencing habits. While the first survey was distributed in summer 2020, the second survey was sent and analyzed in the aftermath of our latest international conference in Broomfield, USA in May 2022. By comparing the answers to the two surveys, which are analyzed in the light of our $CO_2$ emissions assessment, we are in the position to formulate practical recommendations with the goal to establish a transparent strategy to address this intrinsically complex problem shared by scientific communities regardless of their field, scope, size, geography, etc.

**$CO_2$ emissions assessment.** Every third year, the IAS organizes the international conference on the Fundamentals of Adsorption (FOA) – a one week meeting that brings together experts from across the world (**Figure 1**). With approximately 300 participants from more than 30 different countries, it is the most important conference in the field of adsorption and covers all areas from fundamentals to industrial applications. A rotating location with venues in the USA, Europe, and Asia-Pacific is consistent with the international nature of the conference. Participation is only to a small degree dictated by proximity to the venue, and about three quarters of the participants travel from across the world regardless of the location. This results in long (10,000 km) or ultra-long-haul flights (> 10,000 km) for approximately 2/3 of the participants to attend FOA. Such intercontinental travel results in significant $CO_2$ emissions (1-6 t $CO_2$/delegate) that will be difficult to sustain in a net-zero emissions world.

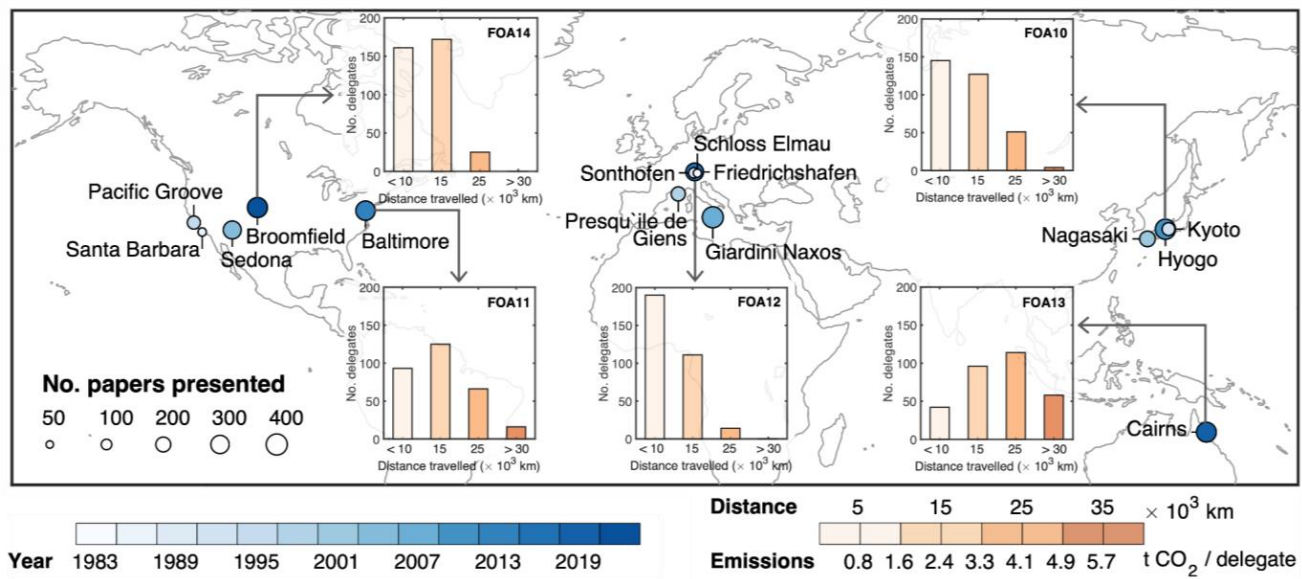

**Figure 1.** World map indicating the location of the International Conference on the Fundamentals of Adsorption (FOA) – the premier international conference in the field of adsorption organized by the IAS. The FOA series of conferences is held every three years, rotating in an alternating manner between the USA, Europe, and the Asian/Pacific area. Since its establishment in 1983, the conference has grown substantially (as quantified by the number of papers presented) and gathers nowadays approximately 300 delegates from 35 different countries. The four insets indicate the average distance traveled by the delegates to reach the conference venue for the last four meetings (FOA 11, Baltimore, USA, 2013; FOA 12, Friedrichshafen at Lake Constance, Germany, 2016; FOA 13, Cairns, Australia, 2019; FOA 14, Broomfield, USA, 2022) and the associated emissions (t $CO_2$/delegate, *see text*). Note that the data for FOA 14 in Broomfield, USA are affected by the lack of attendees from Asia due to COVID-related travel restrictions. On the other hand, for this conference, there was a substantial increase in US attendees with respect to the previous edition in the USA.

In addition to air travel, several other factors add to the $CO_2$ footprint of the conference, albeit to a lesser degree. By selecting FOA 14 as a case study (May 22–27 2022, Broomfield, CO, USA), a detailed analysis was carried out to estimate the average $CO_2$ emissions per participant resulting from the attendance to the conference. The analysis accounts for transportation, accommodation, on-site mobility as well as emissions related to the venue itself, such as those resulting from catering and energy use (**Figure 2**). All details on our $CO_2$ emission assessment can be found in the corresponding section in the Supplementary Information. The resulting cumulative emissions (approx. 830 t $CO_2$ equivalent) amount to an average of 2.6 t $CO_2$ equivalent per participant. To better grasp the reality of this number, a few comparisons can be drawn. In practice, such an amount for a single conference attendance corresponds roughly to half of the world average annual $CO_2$ emissions per person. While this number is already instructive, considering that the typical $CO_2$ emissions per capita scales linearly with the gross domestic product per capita, it is worth acknowledging that the carbon footprint of an individual in a given country is strongly correlated to its economic wealth (with emissions varying from < 1 t $CO_2$ per capita to up to ~ 15 t $CO_2$ per capita). For example, this implies that European residents attending the FOA conference emit in a single week about 1/3 of their average annual emissions. Flights to reach the conference location contribute to 87% of the overall greenhouse gas emissions caused by a scientific conference, whereas accommodation during the conference is responsible for almost two thirds of the remaining 13%. These large carbon emissions suggest that scientific conferences and societies should develop new methods of operation and management to be sustainable in a carbon constrained society. We suggest a threefold approach including emission reduction, compensation of remaining emissions, and approaches to raise awareness.

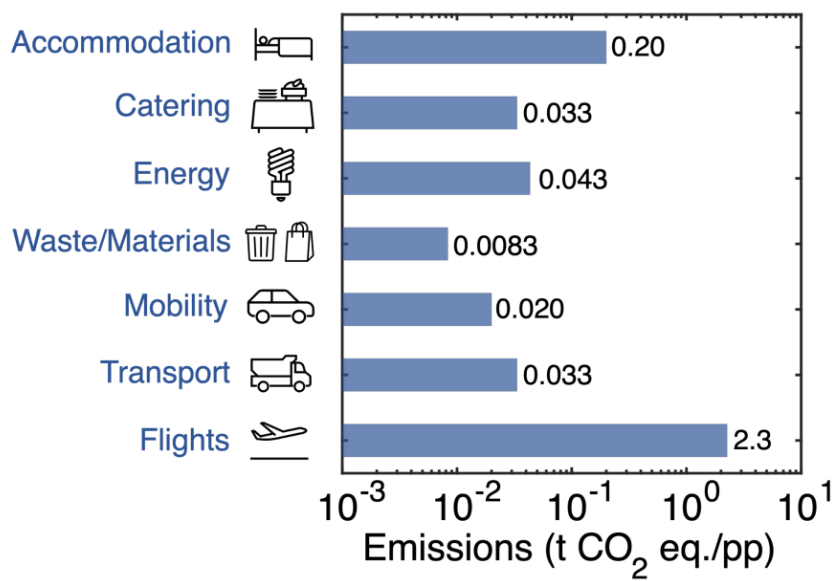

**Figure 2.** Estimated cumulative emissions (t $CO_2$ eq.) associated with the attendance to FOA 14 in Broomfield, CO, USA (May 22–27, 2022). Emissions from flights have been calculated using a detailed flight calculator from myclimate.org,[14] by assuming the same composition of delegates of the previous USA-based conference (FOA 11, Baltimore, MD, USA). Flights were assumed to be all economy and to depart from the capital of each country, or for the US from the respective state capital. For accommodation, the average of a 6-nights hotel stay in Denver was calculated based on hotel footprints.[15] The remaining emissions for onsite mobility, catering, energy consumption, conference materials, waste and transport of goods to/from the venue are based on myclimate.org.[16] All data were cross-checked by considering additional information sources. Additional details on all calculations can be found in the Supplementary Information.

**Pre and post conference surveys**

The first IAS survey was distributed in spring 2020 to the IAS community while the second IAS survey was sent in summer 2022 immediately after the FOA conference in Broomfield, Colorado. For the second survey, an additional section was added to assess how conferencing habits changed between the first and the second survey (pre and during/post COVID times). We also added questions to receive feedback regarding the online attendance experience, which was implemented for the first time for this conference series on the occasion of FOA 14. In practice, online attendance included access to presentation recordings and poster files (available on the first day of the conference and for 3 weeks thereafter) as well as to live stream sessions of the plenary lectures. Finally, for the second survey, questions related to emission estimates and awareness about $CO_2$ compensation were removed. The other questions were repeated from the first survey. The number of participants (~120) and the split between academia and industry (2/3 versus 1/3) were similar in both surveys. More students and postdocs responded to the second survey with an increase from 20% to 32% of the responses from academia, which may be related to the timing of the second survey (directly after the conference) and to announcements made during the conference. For the second survey, 65 % of survey participants attended the conference, 8% attended online, and one third already participated in the first IAS survey from 2020.

For both surveys, after a few general questions regarding their position and knowledge on $CO_2$ emissions related to conference attendance, each survey participant was asked two groups of questions dealing with the following aspects: (1) Carbon footprint reduction through different initiatives such as hybrid conferences combining both on-line and on-site participation, satellite regional (i.e. less impacting) conferences, combined events, etc. together with actions leading to diminished $CO_2$ emissions for on-site physical conference attendance. (2) Use of carbon offsetting actions through reliable organizations ensuring high standards in carbon reduction to compensate for the $CO_2$ emissions caused by the FOA conference. The questions asked in each survey as well as detailed data analysis can be found in the Supplementary Information. When calculating averages and percentages, we only considered participants that actually responded to the question (therefore discarding for a given question survey participants that did not answer).

► *FOA carbon footprint, conference attendance, and satisfaction of participants.* As shown in **Fig. 3(a)**, in both surveys, between 70% and 80% of the survey participants consider it important/very important to reduce FOA carbon footprint (less than ~10% consider it not important/not important at all). Regarding the estimated carbon footprint in the first survey, the answers are broadly distributed with a median of 2 t $CO_2$/participant to attend FOA – a value close to our estimate, as described above. Similarly, on average, the participants believe that flights largely contribute to overall $CO_2$ emissions – with a value close to that estimated from our carbon footprint assessment [87%]. In the second survey distributed about 2.5 years after the COVID outbreak, 58% of respondents indicated that they were planning to attend a similar number of conferences as before the COVID pandemic. Almost half of the participants indicated that they are planning to attend less conferences on site (42%), favor conferences with shorter travel distances (47%), and combine different events (49%). Approximately 58% plan to attend more conferences online than pre-Covid. Importantly, 93% of all survey participants who attended FOA were either satisfied or very satisfied with the social and scientific interactions, however all the online participants were dissatisfied or very dissatisfied with the experience.

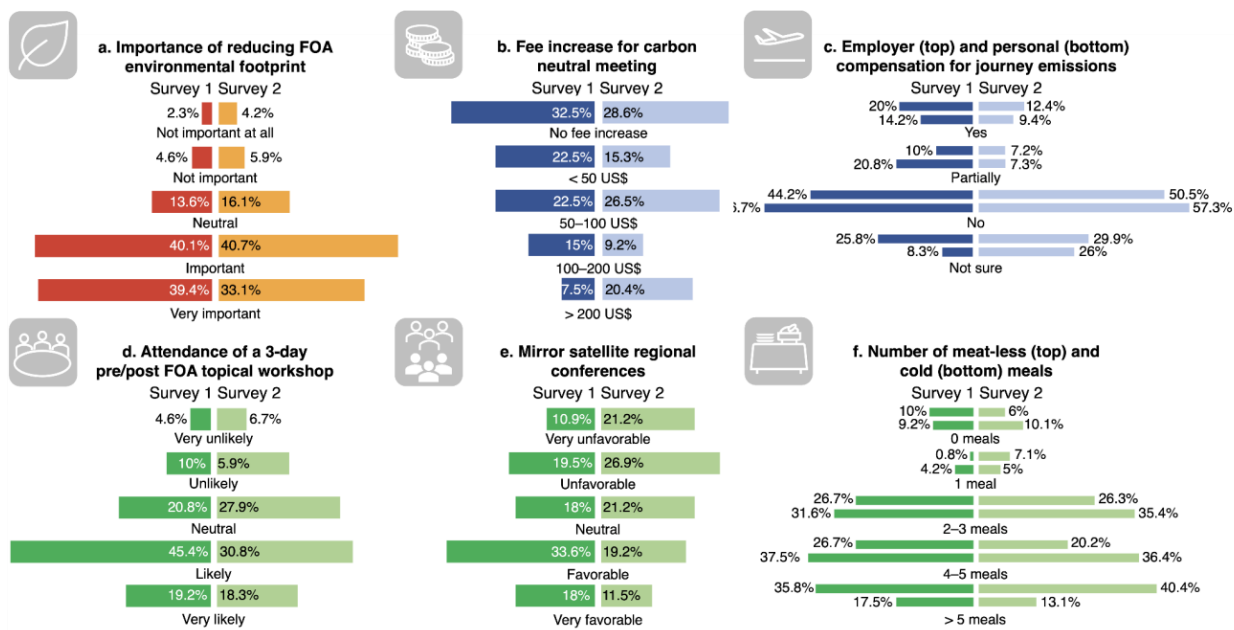

**Figure 3.** Selected results from the two IAS surveys conducted in spring 2020 (Survey 1, dark colored bars on the left-hand side of each histogram) and in summer 2022 (Survey 2, light coloured bars on the right-hand side of each histogram), respectively. Each panel (a to f) refers to a question that was asked in both surveys (with the exact same wording, see Supporting Information). They belong to the three different groups of questions in the surveys, namely *General position and knowledge on $CO_2$ emissions related to conference attendance* (Panel a); *Carbon offsetting initiatives* (Panels b and c); *$CO_2$ emission reduction initiatives* (panels d–f).

► *$CO_2$ emissions reduction.* The survey indicated that 65% of the participants of the first survey are at least somewhat likely to attend a pre/post FOA school/workshop. Similarly, 49% of the participants are likely to very likely to attend such events in the second survey [**Fig. 3(d)**]. The all-inclusive cost judged as acceptable for such events of ~75–175$/day (survey 1: median 125, IQR 75-175; survey 2: median 125, IQR 125-175) remains stable between the two surveys (note that these numbers only consider answers by survey participants, who indicated that they were neutral or likely to attend pre/post conference events). In the second survey, only 31% of participants are (highly) in favor of satellite FOA conferences – conferences that would occur in different places but with the same on-line common sessions – compared to 51% in the first survey. Moreover, 48% of the participants are against this option compared to only 30% in the first survey [**Fig. 3(e)**]. If FOA were broadcasted on-line (on-site attendance combined with possible remote access), most participants indicate that several people around them (including them) would attend online, with a median of 3 (survey 1: IQR 2-5; survey 2: IQR 1-5). As for an acceptable on-line access fee, answers vary broadly from 10 to 400$ with an average of ~100-120$. Interestingly, the number of participants that indicated that nobody would attend such events increased from 9% in the first survey to 20% in the second. In both surveys, the participants supported having several of the seven conference meals served vegetarian (median 4, IQR 3-7) or cold (median 4, IQR 3-5) [**Fig. 3(f)**]. Moreover, ~65 % of participants are (highly) in favor of more informal gala dinners and welcome receptions.

► *Carbon offsetting.* 66% of the participants are aware of carbon compensation by financially supporting emission reduction projects (first survey data, not asked in the second survey). In the first survey, 29% of the participants indicate that their employer is compensating (at least partially) for their flight/trip emissions – this number dropped to 19% in the second survey. A similar decrease was found for compensation on a personal level, with only 17% compensating their personal emission at least partially at the time of the second survey compared to 35% in the first survey [**Fig. 3(c)**]. In both surveys, ~70% of participants support an increase in registration fee to compensate for the $CO_2$ emissions [**Fig. 3(b)**]. In the first survey, when asked about how to apply a registration fee increase to compensate for the carbon footprint [~78$/participant as initially estimated for FOA 14 in Broomfield, CO, USA], approximately half indicated that they prefer such increase to be applied to every participant, with the other half being either against such increase or preferring an optional increase. In the second survey, when answering the same question, ~51% of the participants preferred such an increase to be applied as a general added fee to every participant, while ~28% advocated for an optional fee, and 20% are against. In both surveys, answers about a reasonable fee increase to achieve carbon neutral meetings vary broadly between a few 10$ to a few 100$ for those in favor of a fee increase, resulting in an average of 135$/participant (survey 1, median 100, IQR 50-200) to 220$/participant (median 100, IQR 75-250). When including all participants, values are significantly lower with an average of 91$/participant (survey1) - 130$/participant (mean 50, IQR 0-100 in both cases) [**Fig. 3(b)**].

Overall, the first and second survey data show that reducing their conference-related $CO_2$ footprint is important to the IAS community. Remote conferencing is seen as a good option to achieve this, whereas satellite in-person conferences are perceived as increasingly unattractive. For FOA 14, the conference was experienced as significantly inferior by online attendants compared to onsite attendants (see above for a description of the registration package corresponding to online access). Thus, even though remote conferencing is seen in principle as a good approach for emission reduction, its implementation in practice is often disappointing, and a high quality experience for remote attendants must be ensured to maintain the interest in remote conferencing. Moreover, most participants are in favor of reducing on-site emissions, for example via vegetarian or cold or less fancy meals, shared taxis, less waste, etc., and many comments of survey participants regarding suggestions for making FOA carbon neutral were directed at on-site emission reduction actions. Furthermore, 70% of the attendees are in favor of emission compensation, but only ~50% think that such compensation should be compulsory; despite this, only 10% of the FOA attendees opted out of a $70 CO2 offset fee included in the conference registration. Few people and institutions compensate for their flight/trip emissions. Comments by participants indicate that this is related to their skepticism regarding meaningful compensation actions, and because they favor reduction over compensation. Finding suitable compensation mechanisms at an acceptable price thus is a major challenge and a key bottleneck at the moment for making compensation a viable solution for conference attendees.

**Discussion**

While we are advocating for a shift of paradigm to reinvent conferences as climate neutral, there are important constraints. First, despite all efforts to mitigate the carbon footprint of scientific conferences, the fact that such events are dominated by flight-related greenhouse gas emissions has severe implications. Their climate impact can be reduced by (1) eliminating them or reducing their number, (2) switching to mostly online conferences, or (3) fully compensating the emissions due to air travel associated with conferences. There are obvious drawbacks with all approaches. The scientific community acknowledges that a healthy science requires regular in-person exchange at meetings. The years of anti-Covid measures have made everybody experience how unsatisfactory and ineffective online conferences are. Second, effective emission compensation through real carbon removals (i.e. using negative emission technologies) is not a trivial solution as the associated costs are a moving and increasing target that make them somewhat prohibitive. For instance, the current Compensaid[17] price is ~700 €/t $CO_2$ for immediate compensation through Sustainable Aviation Fuels as compared to the ~30 $/ton $CO_2$ applied on a voluntary basis to the FOA delegates.

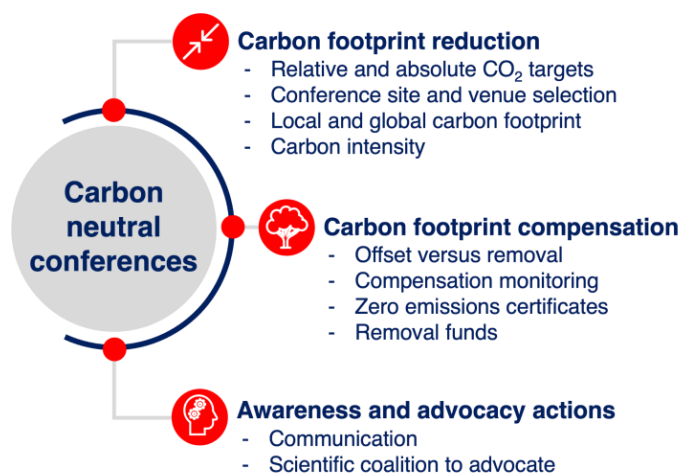

**Figure 4.** Three foundations to enable carbon neutral conferences: reduction, compensation (offset versus removal), and awareness. Details and practical examples on each activity are given in the manuscript text.

In view of these elements, *sobriety in scientific traveling* remains and will remain for quite some time a key component in any realistic strategy to mitigate the impact of conferences on climate. We believe that policy-makers at different levels (e.g. government, institution, university) should enforce carbon neutrality by establishing strong rules monitoring the environmental impact of scientific travel. For instance, it can be argued that reaching carbon neutrality in scientific traveling is at least as important as reducing other $CO_2$ emission sources in research institutions (e.g. buildings, energy consumption, local transportation). In parallel to setting $CO_2$ avoidance targets, a change of mentality in the scientific communities is also mandatory to modify our current mindset about traveling. In particular, one needs to find a better compromise between quantity and quality of the conferences that one attends, without neglecting the vital need for scientists, particularly early career ones, to attend conferences.

Starting from these elements, our working strategy is simple: we intend to provide guidelines for researchers as well as for institutions to design a model for carbon neutral conferences. We consider that scientific conferences are central to the life of science communities so that they deserve to be maintained but in a sustainable format. To reach such an ambitious goal, we propose here a strategy with stringent yet realistic actions that can be implemented without major obstacles (**Fig. 4**). The targeted initiatives listed below are grouped into different categories: (1) carbon footprint reduction, (2) $CO_2$ emissions compensation, and (3) awareness and advocacy actions.

► **Carbon footprint reduction.** Despite their environmental impact, there is a general consensus that scientific conferences should be maintained – albeit in a more carbon neutral form. By fostering vivid exchanges among its members and ensuring the training of the next generation, the majority of participants acknowledges the role of conferences in promoting high quality achievements in a science community. Yet, to maintain conferences, already identified as large $CO_2$ emission contributors, it is also recognized that we should raise the bar in terms of reducing their carbon footprint.

● *Relative and absolute targets.* While the urgency of global warming calls for very stringent targets, we believe that both relative and absolute $CO_2$ objectives should be set. Even if the acute climate crisis is striking everyone on Earth, broad variations among continents and countries are obvious and should be taken into account. In other words, considering local habits and practices, it is difficult – not to say unrealistic – to set the same absolute emissions ($CO_2$ ton/capita) for everybody. However, by setting the same relative target (e.g. share of the national $CO_2$ emissions/capita), we can create a virtuous circle in which every conference organization team will commit to do at least as well as the others.

● *Conference site selection.* We advocate to use $CO_2$ emissions as a stringent selection criterion for conference sites. The specific site location for a conference is often selected for its convenience (e.g. train/plane access) or attractivity (e.g. sightseeing). In the context of global warming, while the organizing team should always be selected for its competence, the conference site selection should also additionally consider carbon footprint. Such criterion, which should be seen as a mandatory point to address in a conference hosting bid, should be formulated to include both local (i.e. on-site) and global $CO_2$ emissions.

● *Local venue.* Universities or research centers are ideal conference venues. In addition to being much cheaper than conference centers, they provide most commodities and facilities for a low carbon footprint. Moreover, such venues offer great video-conferencing possibilities (software, rooms, networks) with IT support. With the numerous students and staff enrolled in universities, they are optimized for conference hosting with sufficient catering and accommodation capacities (universities are also well-equipped for internet access). Organizing events at universities is also virtuous as it fosters the training of future generations. We support that on site access to the conference should be free for all students to trigger vocations – a key aspect to ensure the continuity of any research field. We acknowledge that University-based solutions may only be applicable to small and mid-sized societies.

● *Local carbon footprint.* Carbon footprint at the conference site is far from negligible (~15% of overall emissions). Considering that local emissions are easier to control (compared to flight-related emissions for instance), they should be as close to zero as possible. As discussed below (see *compensation*), we propose to fully remove all local $CO_2$ emissions through financial support to negative emissions solutions. Several measures can be easily taken to reduce the local $CO_2$ emissions, including use of reusable materials, reduced cooling of the venue, shared accommodation, shared taxis, facilitated use of public transport, low-carbon catering including vegetarian food and less sophisticated galas, no printed materials, etc. As an example, the FOA conference organizers utilized a webpage 'Reducing your carbon footprint while at the conference".

● *Global carbon footprint.* The large share of travel-related emissions makes the reduction potential through measures that target other emission sources limited. Thus, as long as air travel is required, any model for carbon neutral conferences should be accompanied by strong policies to decrease the footprint of travel. Since conference organizers do not control travel habits of the participants (which vary broadly), it is in practice unfeasible to leave it to the local team to deal with this enormous problem. However, in any case, for the sake of transparency, all emissions should be estimated as they must be considered when selecting any conference site (so doing will also automatically decrease the overall impact of the organized event).

● *Carbon intensity.* While $CO_2$ emissions are an extensive indicator, carbon intensity is an intensive parameter that describes the impact of an event per day, per capita, etc. Any initiative to decrease carbon intensity does not reduce emissions but makes traveling more worthwhile. For instance, organizing hybrid events with both on-site and on-line attendance reduces carbon intensity at constant (or slightly larger) absolute emissions (for the FOA conference, on-line access led to a ~50% increase in attendance, which reduces the carbon intensity of the FOA meeting when considering both on-site and on-line participants). Similarly, combining events with a pre or post school at the conference site decreases carbon intensity. While receiving less support from the community, other means to decrease carbon intensity include organizing satellite conferences (on different continents at the same time with joint sessions) and reducing the frequency of meetings.

► **Carbon footprint compensation.** Financial support to specific actions to acquire carbon offsets is often invoked to compensate for $CO_2$ emissions. Carbon compensation includes very different strategies: from carbon offsetting obtained by supporting projects through which someone else is avoiding emitting the same amount of $CO_2$ somewhere else, to carbon removals obtained by investing in negative emission technologies. To define an acceptable compensation strategy, we list here potential guidelines.

● *Offset versus removal.* Roughly, carbon offsetting is significantly cheaper than carbon removal (tens versus hundreds US$ per t $CO_2$), but is less permanent. These numbers seem to be increasing rapidly as convenient options – the so-called low hanging fruits – are exploited. It seems fair to assume that the cost difference between offsetting and removal should not change much as the latter solutions are necessarily more complex. Under these

circumstances, it seems unrealistic to remove all $CO_2$ emissions related to a scientific conference (~2.6 ton $CO_2$/capita) without rendering such events unaffordable. In contrast, the following dual mechanism seems to be transparent and sustainable: fully remove all $CO_2$ emissions produced locally at the conference site (after minimizing them as much as possible), while offsetting all travel-related emissions.

● *Compensation monitoring.* To get support from the scientific communities, any carbon compensation must be transparent. We propose to appoint a $CO_2$ fund in each scientific community to monitor and decide any actions performed using the funds raised for compensation. Such panels do not substitute for compensation organizations, whose business is dedicated to such programs, but ensure that high standard carbon offsettings/removals are utilized. Typically, compensation actions should be available to purchase immediately and the associated permanence (i.e. duration) well specified to maximize environmental benefit.

● *Zero emission certificates.* External verification and validation of the carbon compensation initiatives should be sought. This would provide independent, unbiased assessment of the compensation strategy as a part of the overall approach to carbon neutrality. Such evaluation would allow obtaining a zero emission certificate for a given scientific conference. Such labels, which are required to ensure full transparency towards the scientific community and the scientific institutions, would create a virtuous circle by providing reliable numbers on science event-related carbon footprint. In particular, by providing detailed figures about local/global $CO_2$ emissions and carbon compensation actions, such zero emission certificates would serve as a quality label when organizing science events.

● *Compensation funds.* Compensation should be seen as a secondary means to achieve carbon neutrality behind carbon footprint reduction. Yet, considering that zero $CO_2$ emissions are difficult to achieve currently (even locally at the conference site), it is recommended to raise funding for compensation. Optional/mandatory additional registration fees, fundraising through "green" sponsoring, or financial support from institutions can be used to raise funds. Scientists could also request funding for such compensation when applying to funding agencies (we advocate that requested fundings for open access science could be used for carbon compensation instead since preprint servers are available). As an inspiring example, through an optional registration fee (70 US$), the FOA organizers raised 25,000 US$ which exceeded the 23,400 US$ estimated to make this conference carbon neutral through offsetting.

► **Awareness and advocacy.** Actions should be taken to increase the sensitibility of a given community (and beyond) towards the importance of reducing the carbon footprint of science events. All communities should be committed to this fight but some of them must act as role models as their activities are relevant to the climate crisis mitigation. This is the case of the International Adsorption Society whose research field is central to carbon capture and storage; among the many techniques to mitigate $CO_2$ emissions, adsorption processes are mature and viable technologies to remove $CO_2$ from the atmosphere (i.e. negative emission technologies as implemented by e.g.

Climeworks AG, Global Thermostat). Proud of its key-role, the IAS aims at strengthening its visibility and leadership as follows.

● *Communication.* Gaining support from the science communities but also from the society in general is a key objective when fighting the climate crisis. Communication aspects are therefore important to disseminate transparent and unbiased facts to broad audiences; the numerous articles written in specialized and general journals as well as presentations, reports, websites dedicated to the impact of science-related activities are therefore important initiatives. Surveys probing the feedback from either small, well focused communities or broad pans of the society – such as those that were distributed in the pre and post COVID pandemics within the International Adsorption Society membership – also contribute to increasing awareness.

● *Scientific coalition to advocate.* As discussed above, with the current state of the art, a provisional conclusion should be reached: *sustainable science requires climate-neutral aviation*. In view of the need for urgent action to tackle the climate crisis, we argue that scientific societies and academic institutions should not wait passively for the aviation industry to realize carbon neutral operation. We need to take a proactive role beyond that of (for some of us) doing research on new and better solutions. We envision *a scientific coalition of the international scientific societies and academic institutions*, which argue and advocate for climate-neutral flights today, at both the international and the national levels. Such coalitions should serve as science ambassadors on the following topics, among others: (1) climate-neutral aviation, (2) the techno-economic potential and the deployment timeline of sustainable synthetic aviation fuels, (3) the need for carbon-free energy to power their synthesis, (4) the role of $CO_2$ removal solutions to close the carbon cycle associated to flight $CO_2$ emissions, (5) the need of $CO_2$ removal measures to compensate for climate non-$CO_2$ effects and for extra-flight $CO_2$ emissions. Such a coalition should request that standards be established for the calculation of the carbon footprint of everyone's flight, and it should verify that these are consistent with the up-to-date climate science, including about the role of non-$CO_2$ effects. We believe in the power of the scientific community, and in the strengths of its scientific arguments, hence in the role that it can and must play in making aviation climate neutral and research sustainable for the benefit of the society.

**Acknowledgments.** The authors were appointed by the International Adsorption Society (IAS) to form a working group to evaluate its carbon footprint and propose/elaborate mitigation actions. The authors are grateful to the past and current IAS presidents for their trust and support (P. A. Monson, A. Seidel-Morgenstern, and P. Webley). The authors also thank the IAS secretary – Prof. D. Siderius – for his help and all the IAS community for its feedback and support.

**References**
1. Spinellis, D. & Louridas, P. *Plos One* **8**, e66508 (2013).
2. Klöwer, M., Hopkins, D., Allen, M. & Higham, J. *Nature* **583**, 356-359 (2020).
3. Abbott, A. *Nature* **577**, 13 (2020).

4. Jäckle, S. *Political Science & Politics* **54**, 456-461 (2021).

5. Koren, M. & Pető, R. *Plos One* **15**, e0239113 (2020).

6. Seitz, B. M., Aktipis, A., Buss, D. M., Alcock, J., Bloome, P., Gelfand, M., Harris, S., Lieberman, D., Horowitz, B. N., Pinker, S., Wilson, D. S. & Haselton, M. G. *Proc. Nat. Acad. Sci.* **117**, 27767-27776 (2020).

7. Calbi, M., Langiulli, N., Ferroni, F., Montalti, M., Kolesnikov, A., Gallese, V. & Umilta, M. A. *Sci. Rep.* **11,** 2601 (2021).

8. The International Adsorption Society brings together researchers from throughout the world working in industry, academia and government to advance the field of adsorption in areas ranging from the fundamental molecular thermodynamics of adsorption phenomena to the design of industrial separations processes to the applications of adsorption in nanotechnology. https://www.int-ads-soc.org/

9. The IPCC prepares comprehensive Assessment Reports about knowledge on climate change, its causes, potential impacts and response options. https://www.ipcc.ch/reports/

10. Nathans, J. & Sterling, P. *eLife* **5**, e15928 (2016).

11. Jordan, C. J. & Palmer, A. A. *Sci. Adv.* **6**, eabe5810 (2020).

12. Zotova, O., Pétrin-Desrosiers, C., Gopfert, A. & Van Hove, M. www.thelancet.com/planetary-health **4** (2020).

13. Sarabipour, S., Khan, A., Seah, Y. F. S., Mwakilili, A. D., Mumoki, F. N., Sáez, P. J., Schwessinger, B., Debat H. J. & Mestrovic, T. *Nature Human Behaviour* **5**, 296-300 (2021).

14. The myclimate Flight Emission Calculator, published: 13/08/2019, Foundation myclimate https://www.myclimate.org/fileadmin/user_upload/myclimate_-_home/01_Information/01_About_myclimate/09_Calculation_principles/Documents/myclimate-flight-calculator-documentation_EN.pdf

15. https://www.hotelfootprints.org/, accessed: 09/2019

16. https://co2.myclimate.org/en/flight_calculators/new, accessed: 10/2019

17. https://compensaid.com, May 2022. Compensaid is used by such companies as Lufthansa and the price above is realistic from an engineering perspective for high-quality compensation, i.e., the only type of compensation that will survive in the near future.